\documentclass[prl,twocolumn,floatix,amssymb,amsmath,superscriptaddress,showpacs]{revtex4}
\usepackage{hyperref}
\usepackage{graphicx}
\usepackage{color}

\usepackage{amsmath}

\def\one{{\mathchoice {\rm 1\mskip-4mu l} {\rm 1\mskip-4mu l} {\rm \mskip-4.5mu l} {\rm 1\mskip-5mu l}}}
\def\ket#1{|#1\rangle}
\def\bra#1{\langle#1|}

\def\topfraction{0.9}
\def\bottomfraction{0.9}
\def\intextfraction{0.5}
\def\textfraction{0.1}

\begin{document}

\title{Coherent control of two nuclear spins using the anisotropic hyperfine interaction}

\author{Yingjie Zhang}
\affiliation{Institute for Quantum Computing, University of Waterloo, Waterloo, Ontario, N2L 3G1, Canada}
\affiliation{Department of Physics and Astronomy, University of Waterloo, Waterloo, Ontario, N2L 3G1, Canada} 

\author{Colm A. Ryan}
\affiliation{Institute for Quantum Computing, University of Waterloo, Waterloo, Ontario, N2L 3G1, Canada}
\affiliation{Department of Physics and Astronomy, University of Waterloo, Waterloo, Ontario, N2L 3G1, Canada}

\author{Raymond Laflamme}
\affiliation{Institute for Quantum Computing, University of Waterloo, Waterloo, Ontario, N2L 3G1, Canada} 
\affiliation{Department of Physics and Astronomy, University of Waterloo, Waterloo, Ontario, N2L 3G1, Canada}
\affiliation{Perimeter Institute for Theoretical Physics, Waterloo, Ontario, N2J 2W9, Canada}

\author{Jonathan Baugh\footnote{corresponding author: baugh@iqc.ca}}
\affiliation{Institute for Quantum Computing, University of Waterloo, Waterloo, Ontario, N2L 3G1, Canada}
\affiliation{Department of Physics and Astronomy, University of Waterloo, Waterloo, Ontario, N2L 3G1, Canada} 
\affiliation{Department of Chemistry, University of Waterloo, Waterloo, Ontario, N2L 3G1, Canada}

\date{\today}


\begin{abstract}
We demonstrate coherent control of two nuclear spins mediated by the magnetic resonance of a hyperfine-coupled electron spin. This control is used to create a double nuclear coherence in one of the two electron spin manifolds, starting from an initial thermal state, in direct analogy to the creation of an entangled (Bell) state from an initially pure unentangled state. We identify challenges and potential solutions to obtaining experimental gate fidelities useful for quantum information processing in this type of system. 
\end{abstract}
\maketitle
\section{Introduction}
\indent Solid-state spin systems are interesting candidates for quantum information processing: the small systems explored in the lab today are excellent test-beds for the ideas of quantum control and quantum error correction, and it may be possible to reach enough qubits for non-trivial quantum computations, or to integrate these systems into useful hybrid devices for quantum communications \cite{PhysRevLett.96.070504} or quantum sensors \cite{Maze:2008kx}. The past two decades have seen much progress in the high fidelity control of small quantum processors realized by nuclear magnetic resonance (NMR) \cite{Baugh2007, Cory2000}, electron spin resonance (ESR) \cite{Hodges:2008ys, PhysRevLett.92.076401, Du2009, Mitrikas2010, PhysRevLett.105.200402} and electron-nuclear double resonance (ENDOR) \cite{Mehring:2003rt, scherer:052305,Simmons2011,neumann2008}. These experiments have served as benchmarks for experimentally attainable gate fidelities \cite{Ryan:2009rt}, and have spurred the development of robust quantum control methods \cite{PhysRevA.75.012322, Nigmatullin2009}. Hybrid electron-nuclear spin systems make it possible to exploit the strengths of each type of spin: electron spin for initialization, readout and control, and nuclear spin for long storage and coherence times\cite{neumann2008, jiang2009, Simmons2011}. In particular, it is advantageous to use the electron spin as an actuator to gain full control of the system's spin dynamics via the anisotropic part of the hyperfine interaction \cite{Hodges:2008ys,Khaneja2007}. Since the hyperfine interaction to nearby nuclei can be of order $1-100$ MHz, fast electron-nuclear and nuclear-nuclear gates can be realized by this approach. Several prior studies have demonstrated coherent control of a one electron + one nuclear spin system using a modulated microwave field in concert with anisotropic coupling \cite{Hodges:2008ys, Du2009, Mitrikas2010}. In this work, we demonstrate an entangling gate between two nuclear spins fully mediated by control of the electron spin. This is an important first step towards achieving efficient control of hybrid electron-nuclear spin systems of interest for quantum information processing. \\
\section{Experiment}
\indent The spin system employed here is based on the stable radical of malonic acid in the solid state \cite{Mehring:2003rt, Hodges:2008ys, Du2009, Mitrikas2010} with an additional $^{13}$C labeling. X-ray irradiation removes a proton from the methylene group leaving behind an unpaired $\pi$ electron. The electron spin has g-factor $g = 1.9843$ and couples via the Fermi contact and dipolar hyperfine interactions to the remaining methylene proton and to the $^{13}$C-labeled methylene carbon. The tensors describing the $^{1}$H and $^{13}$C hyperfine interactions were reported in ref. \cite{Cole1961}. Pulsed electron spin resonance was performed at an X-band microwave frequency of $9.1875$ GHz at room temperature on a home-built spectrometer. Numerically-derived optimal control pulses were partially corrected for the finite resonator bandwidth and other pulse imperfections by installing a pick-up antenna near the resonator and adjusting the input pulse until the measured pulse best matched the desired waveform. Additional experimental details may be found in the supplementary material \cite{som}. \\
\begin{figure*}[t]
\includegraphics[width= 15cm]{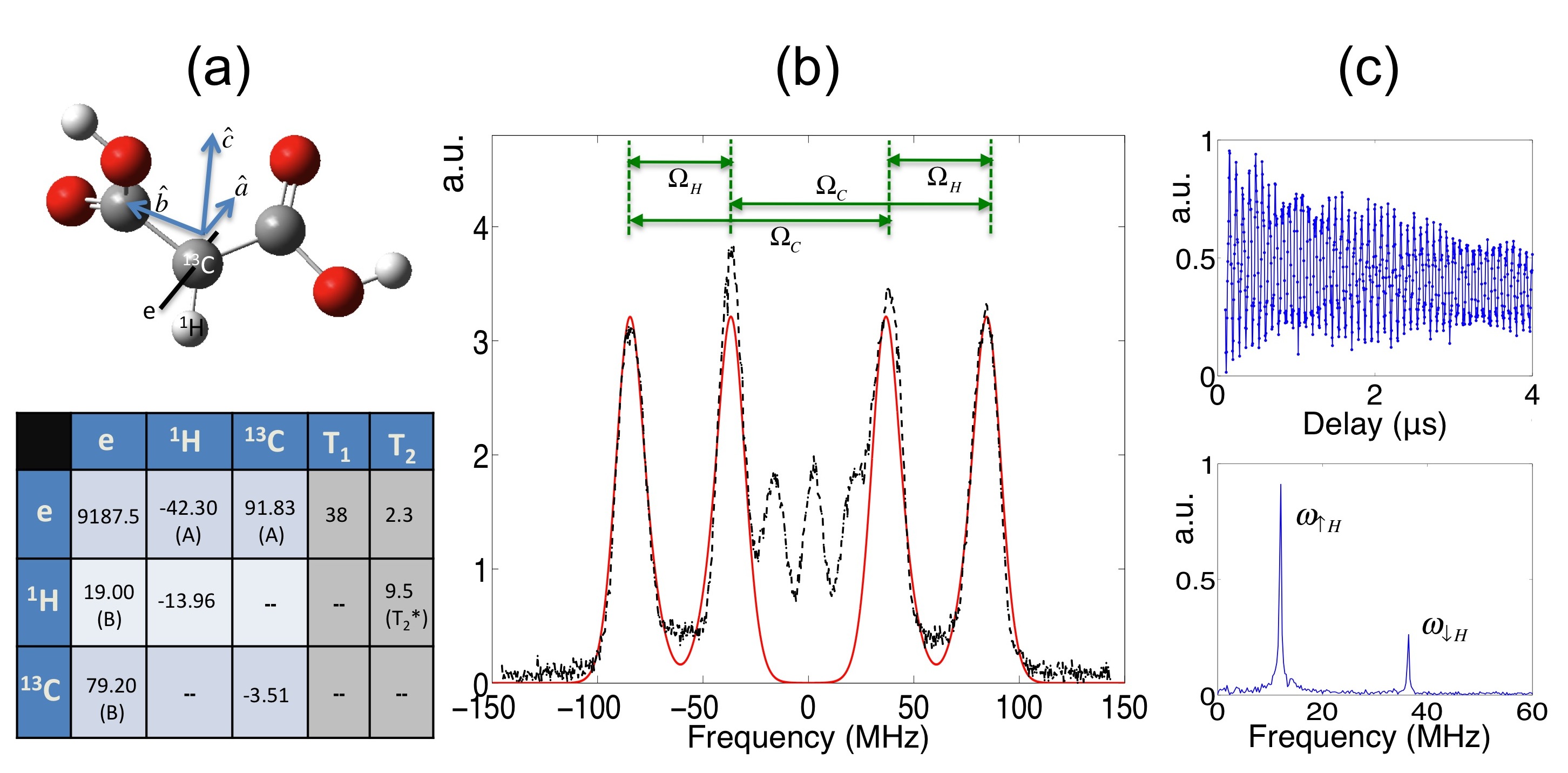} 
\caption{(a, upper) Schematic of the electron-$^{13}$C-$^{1}$H system on the malonic acid radical. The electron occupies a $\pi$-orbital oriented along the $\hat{a}$-axis, perpendicular to the $^{13}$C-$^{1}$H bond along the $\hat{c}$-axis of the Cartesian $(a,b,c)$ coordinate system shown ($\hat{b}$, $\hat{c}$ are in the plane of the three carbon atoms). (a, lower) Table of Hamiltonian parameters for the crystal orientation used in this work, with external magnetic field direction $\hat{B_0} = (-0.55,0.27,0.79)$. Larmor frequencies appear along the diagonal and hyperfine coupling coefficients are off-diagonal, with all frequencies in MHz. The two columns at right list observed $T_1$ and $T_2$ relaxation times, in microseconds. (b) Fieldswept ESR spectrum with experimental data (dotted, black line) and best-fit simulation (solid, red line). Additional peaks near the central region of the spectrum are due to another defect present in the crystal, and may be ignored. The $^{13}$C and $^{1}$H splittings are $\Omega_C = 122$ MHz and $\Omega_H = 48$ MHz, respectively. (c) Upper panel shows three-pulse ESEEM data recorded with hard pulses resonant with the transition at $+37$ MHz, and in the lower panel its Fourier transform shows the two $^{1}$H transition frequencies $\omega_{\uparrow H} = 11.99 \pm 0.01$ MHz and $\omega_{\downarrow H} = 36.35 \pm 0.04$ MHz corresponding to electron spin-up and spin-down states, respectively.} \label{fig1}
\end{figure*}
The secular internal spin Hamiltonian of the one electron + two nuclear spin system is given by
\begin{equation}
\mathcal{H} = \Omega_e S_z + \omega_H I^H_z + \omega_C I^C_z + S_z \otimes  \sum_{k\in{C,H}} (A_k I^k_z + B_k I^k_x) 
\end{equation}
with the component of electron spin $S_z$ along the external field direction $\hat{z}$, nuclear spin operators $\textbf{I}^k$, and where $\Omega_e$, $\omega_H$ and $\omega_C$ are the electron, $^{1}$H and $^{13}$C Zeeman frequencies, respectively, and $\{A_k, B_k\}$ are the four hyperfine coefficients. The small nuclear-nuclear dipolar coupling is neglected. In this system $\Omega_e >> A_k, B_k > \omega_H > \omega_C $, and the primary orientation dependence of the Hamiltonian is due to the hyperfine coefficients. Each nuclear spin has an anisotropic term $B S_z I_x$ that couples the longitudinal component of electron spin to a transverse component of nuclear spin. In combination with the nuclear Zeeman term, this causes the nuclear spin quantization axes to be dependent on the electron spin state, and in general to be non-collinear with the external magnetic field (details on exploiting the anisotropic part of the hyperfine interaction are given in the supplementary material \cite{som}). We may choose logical qubit states of the nuclear two-spin system to be the energy eigenstates in either of the two manifolds defined by the electron spin up or down; for example, the logical $^{1}$H qubit states in the electron $\ket{\uparrow}$ manifold are
\begin{align}
\ket{0_H} &= \cos{(\theta^{H}_{\uparrow}/2)}\ket{\uparrow_H} - \sin{(\theta^{H}_{\uparrow}/2)}\ket{\downarrow_H}\\
\ket{1_H} &= \sin{(\theta^{H}_{\uparrow}/2)}\ket{\uparrow_H} + \cos{(\theta^{H}_{\uparrow}/2)}\ket{\downarrow_H}
\end{align}
where $\ket{\uparrow_H}$ and $\ket{\downarrow_H}$ are the nuclear eigenstates of the Zeeman interaction, and $\theta^{H}_{\uparrow} = \tan^{-1}\left(\frac{-B_H}{2\omega_H + A_H} \right)$ is the angle between the proton quantization axis and the external field direction when the electron spin is up. Analogous expressions apply for the carbon spin states with angle $\theta^{C}_{\uparrow}$, and for the electron spin down manifold with angles $\theta^{H,C}_{\downarrow} =  \tan^{-1}\left(\frac{-B_{H,C}}{-2\omega_{H,C} + A_{H,C}} \right)$. An allowed ESR transition, i.e. one that does not change the nuclear spin states in the limit of vanishing anisotropic terms, can be driven at a rate $\omega_{1}\cos{\left(\frac{\theta_{\uparrow}-\theta_{\downarrow}}{2}\right)}$ by a resonant microwave field of amplitude $\omega_{1}$. A forbidden transition, i.e. one involving one ore more nuclear spin flips, is driven at a rate $\omega_{1}\sin{\left(\frac{\theta_{\uparrow}-\theta_{\downarrow}}{2}\right)}$. In our system, the forbidden transitions involving $^{13}$C spin flips are strongly suppressed relative to the $^{1}$H transitions, since the $^{13}$C Zeeman frequency is very small in comparison with its secular hyperfine coefficient, $|A_C|>>|2\omega_C|$. The Hamiltonian parameters obtained for the crystal orientation used in these experiments are listed in the Table of Fig. 1a.\\
\indent For the crystal orientation we used, the fieldswept ESR spectrum and the three-pulse ESEEM data are shown in Figures 1(b) and 1(c), respectively. Note that only the $^{1}$H transition frequencies appear in the ESEEM measurement due to the strong suppression of the forbidden transitions that involve the $^{13}$C nucleus. The ESEEM spectra provide about two orders of magnitude better frequency precision than the fieldswept due to the much longer nuclear $T^*_2$ timescales. The $^{1}$H ESEEM frequencies are combined with the best fit frequencies to the fieldswept data, and together with the known $^{1}$H and $^{13}$C hyperfine tensors, a most likely spin Hamiltonian is obtained. With this estimated Hamiltonian, optimal control pulses are numerically derived to perform the experiment described below; this experiment is equivalent to a `targeted' ESEEM that excites a double nuclear coherence, giving us direct information about the $^{13}$C Hamiltonian parameters and further refining the Hamiltonian estimate \footnote{The initial Hamiltonian estimate was found to agree with the refined estimate to within $1.6\%$, in the sense that $\|\mathcal{H} - \mathcal{H'}\| < 0.016\|\mathcal{H}\|$, where $\mathcal{H}$ and $\mathcal{H'}$ are the original and refined estimates for the rotating frame Hamiltonian.}.\\
\indent To demonstrate coherent control of the nuclear spin states, we implement an entangling quantum gate, for example the operation that maps $\ket{01}\rightarrow \frac{\ket{01}\pm\ket{10}}{\sqrt{2}}$. The experiment is laid out schematically in figure 2(b). Starting from the thermal state deviation density matrix $\rho_{0} = S_z$, a selective $\pi$ pulse is applied to invert the population of the transition at $\Omega_e + 37$MHz, followed by a 0.8 $\mu$s optimal control GRAPE \cite{PhysRevA.75.012322} pulse designed to perform a $\pi/2$ rotation in the submanifold of two nuclear states labeled in figure 2(a) by the 22 MHz transition; the coherence between these two states is a double nuclear coherence that behaves analogously to a pure state $\frac{\ket{01}+e^{i\phi}\ket{10}}{\sqrt{2}}$. Note that the latter operation could also be accomplished with an additional radiofrequency (RF) channel to directly drive the nuclear transition (ENDOR), but would require at least several microseconds with typically available RF powers, as well as the complication of an additional RF interface. Subsequent to the selective $\pi$ (inversion) pulse, the ideal state of the system is 
\begin{widetext}
\begin{equation}
\rho = -\ket{\downarrow}\bra{\downarrow} \otimes (\textbf{E}^H_0\otimes \textbf{E}^C_0 + \textbf{E}^H_1\otimes \textbf{E}^C_0 -\textbf{E}^H_0\otimes \textbf{E}^C_1 +\textbf{E}^H_1\otimes \textbf{E}^C_1)+\ket{\uparrow}\bra{\uparrow} \otimes (\textbf{E}^H_{0'}\otimes \textbf{E}^C_{0'} + \textbf{E}^H_{1'}\otimes \textbf{E}^C_{0'} -\textbf{E}^H_{0'}\otimes \textbf{E}^C_{1'} +\textbf{E}^H_{1'}\otimes \textbf{E}^C_{1'})
\end{equation}
\end{widetext}
where $\textbf{E}^j_m = \ket{m}\bra{m}$ is the density matrix corresponding to the energy eigenstate $\ket{m}$ ($m\in{0,1}$) for nuclear spin $j$, and we label the eigenstates in the spin-up manifold by $\ket{m'}$. The GRAPE pulse performs the following transformation in the spin-down manifold:
\begin{equation}
\textbf{E}^H_1\otimes \textbf{E}^C_0 -\textbf{E}^H_0\otimes \textbf{E}^C_1 \Rightarrow  \ket{-}\bra{-} - \ket{+}\bra{+}
\label{test}
\end{equation}
where $\ket{\pm} = \frac{\ket{0_H 1_C} \pm \ket{1_H 0_C}}{\sqrt{2}}$, and the pulse acts as the identity operator on all other terms. The terms on the right side of Eq. ~(\ref{test}) evolve during a free evolution period $\tau$ as
\begin{align}
&\cos{\left((\omega_{\downarrow C} - \omega_{\downarrow H})\tau\right)}\left(\ket{-}\bra{-} - \ket{+}\bra{+}\right) \nonumber \\
&+ \sin{\left((\omega_{\downarrow C} - \omega_{\downarrow H})\tau\right)}\left(\ket{+}\bra{-} + \ket{-}\bra{+}\right)
\end{align}
which is equivalent to precession of the pure state $\frac{\ket{0_H 1_C}+e^{-i\tau(\omega_{\downarrow C} - \omega_{\downarrow H})}\ket{1_H 0_C}}{\sqrt{2}}$ due to the internal Hamiltonian. Applying the GRAPE pulse again reverses the transformation in Eq. ~(\ref{test}), so that the diagonal terms in the electron spin down manifold ($\ket{\downarrow}\bra{\downarrow}$) become
\begin{align}
\textbf{E}^H_0\otimes \textbf{E}^C_0 &+\cos{\left((\omega_{\downarrow C} - \omega_{\downarrow H})\tau\right)}(\textbf{E}^H_1\otimes \textbf{E}^C_0 -\textbf{E}^H_0\otimes \textbf{E}^C_1) \nonumber \\
&+\textbf{E}^H_1\otimes \textbf{E}^C_1
\end{align}
To perform readout, a selective $\pi/2$ pulse is applied on the electron transition corresponding to the nuclear spin term $\textbf{E}^H_0\otimes \textbf{E}^C_1$. The observable is the $\hat{z}$-electron spin magnetization on this transition, and it can be shown \cite{som} that the normalized signal as a function of delay $\tau$ is simply
\begin{equation}
S(\tau)  =  \frac{1+ \cos{\left((\omega_{\downarrow C} - \omega_{\downarrow H})\tau\right)}}{2}
\end{equation}
\begin{figure}[ht]
\includegraphics[width= 8cm]{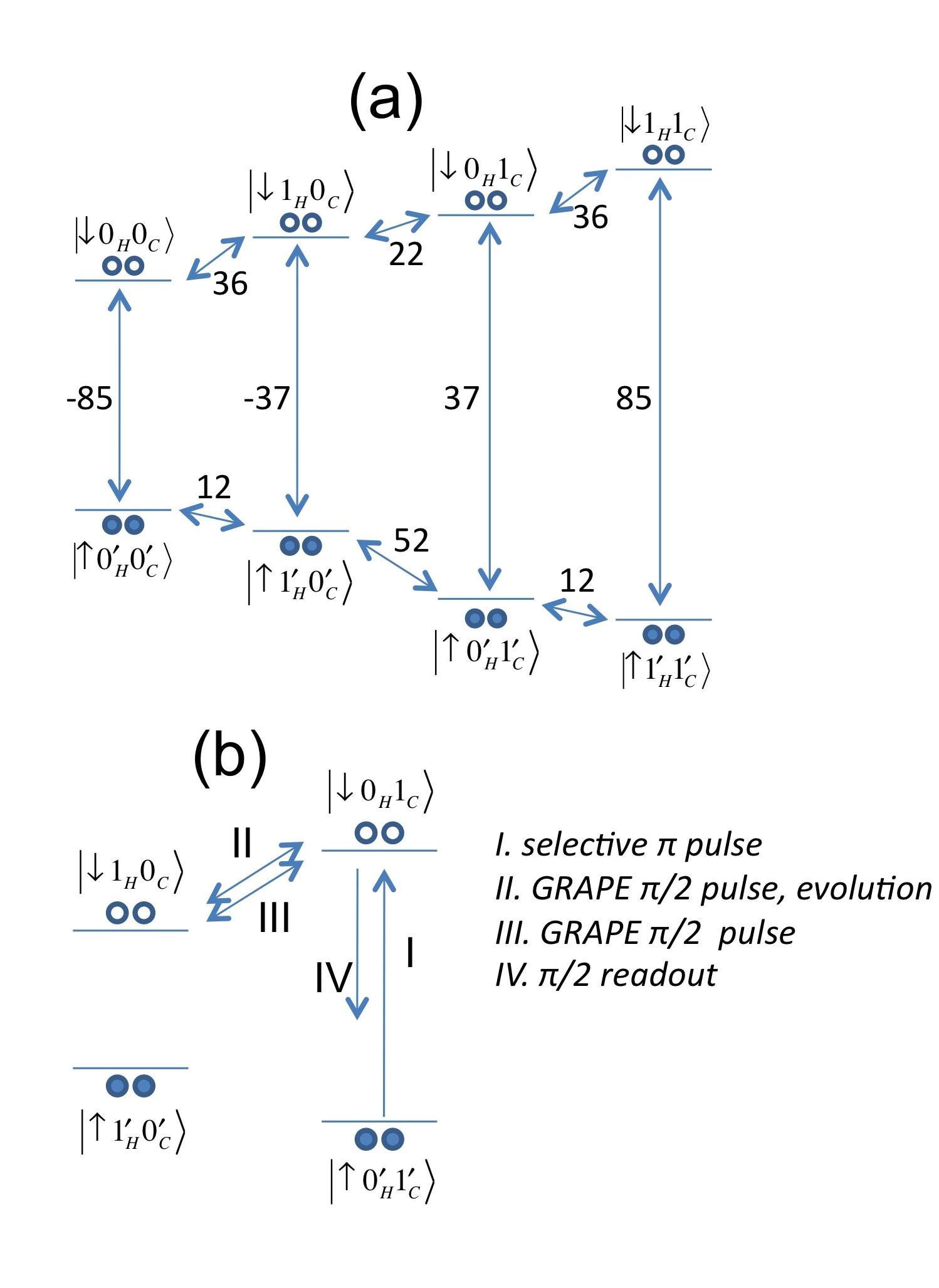} 
\caption{(a) Energy level diagram for the 3-spin system showing transition frequencies (in MHz) between eigenstates. $\ket{\uparrow}$ and $\ket{\downarrow}$ label the electron manifolds. Note that the nuclear quantization axes depend on the electron spin state, so that nuclear eigenstates in the two electron spin manifolds are different. (b) Schematic of the experimental sequence for creating and detecting double nuclear coherence in the electron spin-down manifold. Steps 1 and 4 refer to driving direct electronic transitions, while steps 2 and 3 involve an operation similar to a Hadamard gate (labeled `GRAPE $\pi/2$ pulse') between the two nuclear sublevels $\ket{01}$ and $\ket{10}$. } \label{fig2}
\end{figure}
\indent The GRAPE pulse used in this experiment had a unitary fidelity of $98\%$ on the ideal system, i.e. without including ensemble inhomogeneity effects (inhomogeneities of the DC and microwave magnetic fields throughout the sample), electron spin dephasing ($T_{2e}$), or other error sources such as the finite bandwidth of the ESR resonator and pulse imperfections \footnote{Note, however, that numerical optimization of the GRAPE pulse was performed over a discrete set of DC fields designed to roughly match the $T^*_{2e}$ linewidth in order to improve the `robustness' of the pulse. Without this constraint, much higher fidelity pulses on the ideal system are easily found, but these pulses would perform much worse in the actual experiment.}. When realistic $T_{2e}$ and DC field inhomogeneity ($T^*_{2e}$ linewidth) are included in the simulation, the fidelity drops to $68\%$, with each factor contributing roughly equal amounts to the error (note that the GRAPE pulse duration is about $1/3$ of $T_{2e}$ in the present system). Spatial inhomogeneity of the microwave field and other pulse imperfections reduce the actual fidelities even further. \\
\indent The experimental results are summarized in figure ~\ref{fig3}. Fig. ~\ref{fig3}a shows the 22 MHz modulation of the readout echo signal due to the evolution of the double nuclear coherence in the electron spin-down manifold. Simulated data are also presented for comparison (blue dashed line); these were calculated at the same discrete time points as the real data, and include the electron $T_2$ as well as the ensemble inhomogeneity ($T^*_2$) effects. The Fourier transforms of the simulated and experimental data are both strongly peaked at 22 MHz as expected, shown in the left panel of (b). The experiment was also carried out on the 52 MHz transition in the electron spin-up manifold using a different GRAPE pulse, with results shown in the right panel of (b). For both experiments, the modulation amplitude of the measured signal is about 1.6 times smaller than that predicted by simulation. The best fits to the experimental data give frequencies 22.3 $\pm$ 0.5 MHz and 52.0 $\pm$ 0.4 MHz in excellent agreement with the estimated Hamiltonian. Note that a small amount of 12 MHz modulation due to the $^{1}$H transitions in the spin-up manifold is present in the 22 MHz transition data, both in simulation and experiment. We emphasize that we have not been able to reproduce the 22 MHz and 52 MHz frequencies in any type of standard ESEEM experiment, including `matched' ESEEM \cite{Jeschke1998} with soft pulses designed to favor excitation of those coherences.\\
\begin{figure}[t]
\includegraphics[width= \columnwidth]{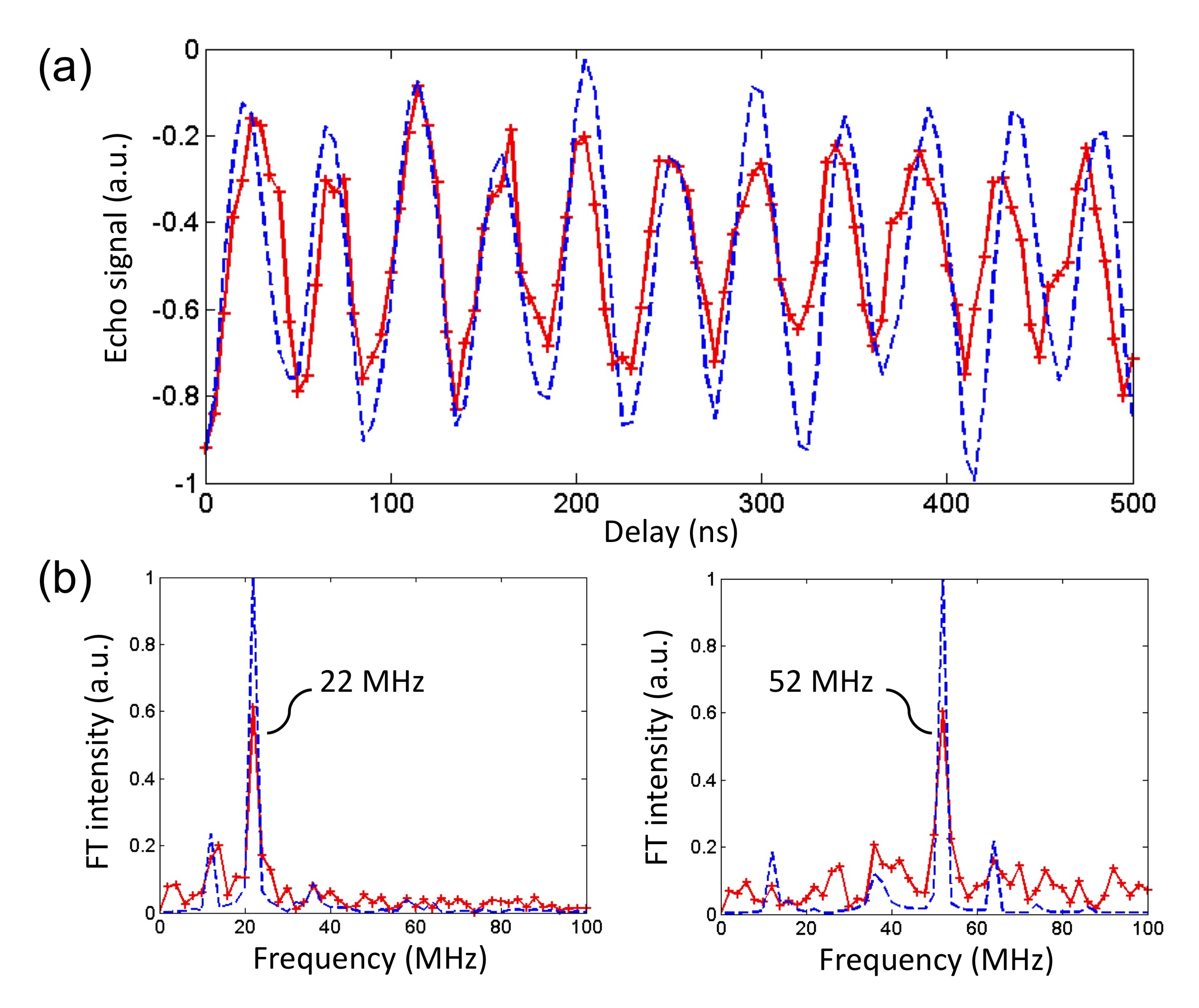} 
\caption{(a) The double-nuclear coherence echo signal versus delay time in experiment (red crosses, solid line) and simulation (blue dashed line) for the 22 MHz transition in the electron spin-down manifold. (b) The Fourier transform of the oscillatory signal for the 22 MHz transition experiment (left), and for the analogous experiment performed on the 52 MHz transition in the electron spin-up manifold (right).} \label{fig3}
\end{figure}
\section{Conclusion}
In summary, we have demonstrated experimentally that an electron can be used as an actuator to perform an entangling gate between two nuclear spins. This is a first step towards reaching high-fidelity, universal control of one electron + $N$-nuclear spin hyperfine coupled systems using a single microwave field. Several factors limited the fidelity of coherent control achievable in this experiment. First, the bandwidth of the loop-gap resonator must cover the full spectral width so that all the system transitions can be driven efficiently. In our experiment, the $Q$-factor could only be spoiled to achieve a bandwidth $\approx 140$ MHz, smaller than the full spectral width of $170$ MHz; indeed, some features of the GRAPE pulses involving high frequency components  could not be adequately corrected with our pickup antenna feedback method \cite{som}. Secondly, spoiling the $Q$-factor reduces the maximum Rabi frequency proportionately; when the Rabi frequency $\nu_{Rabi} << \|\mathcal{H}_{rot}\|$ where $\mathcal{H}_{rot}$ is the rotating-frame Hamiltonian, the system is still controllable in principle but requires pulses that are significantly longer than the nearly time-optimal pulses that can be found when $\nu_{Rabi} \gtrsim \|\mathcal{H}_{rot}\|$. When the latter condition is satisfied, we find empirically that pulses with much better robustness to ensemble inhomogeneities can be obtained, and furthermore, efficient decoupling of the electron from the nuclear spin bath would be possible with hard pulses, allowing $T_{2e}$ to be significantly lengthened  \cite{Du2009,PhysRevLett.105.200402}. The coherence times of the nuclei as well as the electron ultimately depend on the relaxation time $T_{1e}$ of the electron, which dramatically increases at low temperatures ($T<<10K$)  in malonic acid \cite{Dalton197277} and in many other spin systems. We are presently working to extend these experiments to that regime, and to utilize spin systems with narrower spectral widths. Finally, we remark that simulations support the theoretical result \cite{Hodges:2008ys, Khaneja2007} that it is possible to construct any desired unitary operator in the Hilbert space of the 3-spin system. For example, control-NOT and SWAP gates acting on the two nuclei can be found numerically with similar durations and unitary fidelities as the pulse used in the experiments above, with the same constraints. \\

\textbf{Acknowledgements --} 
We would like to acknowledge fruitful discussions with D. G. Cory and technical assistance from J. Chamilliard, M. Ditty and H. van der Heide. This work was supported by the Natural Sciences and Engineering Research Council of Canada and the Canada Foundation for Innovation. 
\bibliography{esr}

\newpage
\appendix

\def\one{{\mathchoice {\rm 1\mskip-4mu l} {\rm 1\mskip-4mu l} {\rm \mskip-4.5mu l} {\rm 1\mskip-5mu l}}}
\def\ket#1{|#1\rangle}
\def\bra#1{\langle#1|}

\def\topfraction{0.9}
\def\bottomfraction{0.9}
\def\intextfraction{0.5}
\def\textfraction{0.1}

\vfuzz2pt 

\section{Supplementary material for `Coherent control of two nuclear spins using the anisotropic hyperfine interaction'}

\subsection{Exploiting the anisotropic hyperfine interaction}
\indent To understand the control of nuclear spin states mediated by the anisotropic part of the hyperfine interaction, we need only consider one nuclear spin-1/2 coupled to an electron spin in an external magnetic field. To a very good approximation the system with nonzero anisotropic coupling is described by an internal Hamiltonian \cite{schweiger2001}
\begin{equation}
\mathcal{H}_{0} = \Omega_e S_z + \omega_n I_z + AS_zI_z + B_x S_z I_x + B_y S_z I_y
\end{equation}
where the couplings are `secular' with respect to the electronic operator $S_z$ because $|\Omega_e|>>|A|, |B|$, but the $B$ terms are nonsecular with respect to the nuclear Zeeman term (`pseudosecular') since typically $|\omega_n| \lesssim |B|$. The pseudosecular terms can be unitarily transformed to a single term $ B S_z I_x$ without loss of generality. For an axially symmetric hyperfine tensor, which is the case for both the $^{13}$C and $^{1}$H tensors in the malonic acid system studied here, the coefficients are given by  \cite{schweiger2001}
\begin{align}
A &= A_{iso} + D(3\cos^2{\theta} - 1)\\
B &= 3D\cos(\theta)\sin(\theta)
\end{align}
where $A_{iso}$ is the isotropic term due to the Fermi contact interaction, $D$ is the dipolar coupling between the electron and nuclear spins, and $\theta$ is the angle of the external magnetic field with respect to the symmetry axis of the tensor. It is the pseudosecular term $B S_z I_x$ that allows for universal control of the nuclear spin state when there is an additional control term $\omega_{rf}(t) S_x$ resonant with the electron transitions present \cite{Hodges:2008ys,Khaneja2007}. Therefore, it is critical to orient the external magnetic field so that $B$ is non-zero and as large as possible for each of the nuclear spins, whose hyperfine principle axis orientations are in general different. It is also clearly important that the external magnetic field be not so strong that $|\omega_n| >> |B|$, otherwise the pseudosecular term is suppressed and universal control is not obtained. \\
\indent The  pseudosecular term $BS_zI_x$ allows for universal control because it creates non-parallel quantization axes in the nuclear spin subspace that depend on the state of the electron. When the electron is in the $m_s=-\frac{1}{2}$ state, the nuclear quantization axis is $-\frac{B}{2}I_x + (\omega_n -\frac{A}{2})I_z$, whereas in the  $m_s=+\frac{1}{2}$ state, the axis lies along $+\frac{B}{2}I_x + (\omega_n +\frac{A}{2})I_z$.  Thus by modulating the electron spin state, rotations may be performed about two non-commuting axes, providing universal control of the nuclear spin. An illustrative example can be found in \cite{Mitrikas2010}. \\
\subsection{Additional experimental details}
\indent The single crystal of malonic acid was approximately 4mm $\times$ 1mm $\times$ 1mm in size and was mounted inside an Al loop gap resonator \cite{Hyde1989} with an intentionally low quality factor $Q\approx 65$ to achieve high bandwidth control. Numerically derived optimal control pulses \cite{Khaneja2005296} were synthesized on a Tektronix arbitrary waveform generator with minimum timing resolution 1ns, and amplified with a 500W traveling wave tube amplifier. The maximum Rabi frequency was limited to 28 MHz due to the spoiled $Q$-factor. \\
\indent Figure 1(b) in the main text shows the fieldswept ESR spectrum obtained via spin echo. The crystal orientation was chosen as a compromise between having well resolved ESR transitions together with reasonably large anisotropic coefficients, while limiting the total spectral width to match the experimentally available control bandwidth. The spectral width $\Omega_C + \Omega_H$ is dominated by the carbon splitting $\Omega_C = 122$ MHz. The inhomogeneous broadening of the ESR transitions is mainly due to dipolar coupling with the OH protons and protons on surrounding molecules in the crystal; from the locations of protons in the crystal structure we have estimated this dipolar linewidth to be 14 MHz in good agreement with the measured linewidth, and corresponding to a $T^*_{2e} = 23$ ns. The electron single-echo decay time is $T_{2e} \approx 2.3 \mu$s. Figure 1(c) in the main text shows the signal obtained under a standard three-pulse ESEEM sequence versus the delay between second and third pulses. This sequence transfers coherence to nuclear spin transitions prior to the delay, so that the decay rate corresponds to the nuclear $T^*_2$. The Fourier transform of the time series data yields the nuclear transition frequencies. \\
\indent Due to the strong suppression of the $^{13}$C forbidden transitions, three-pulse ESEEM is only able to detect $^{1}$H transition frequencies, shown in the lower panel of Fig. 1(c) (main text). For example, the forbidden transition $\ket{\uparrow 0'_H 1'_C} \Leftrightarrow \ket{\downarrow 1_H 0_C}$ is related to the creation of a double nuclear coherence, and it is suppressed relative to the allowed transitions by a factor $\left|\tan{\left(\frac{\theta^H_{\uparrow}-\theta^H_{\downarrow}}{2}\right)}\tan{\left(\frac{\theta^C_{\uparrow}-\theta^C_{\downarrow}}{2})\right)}\right| \approx 0.013 $, made small by the $^{13}$C contribution. The two transition frequencies observed in ESEEM correspond to the $^{1}$H transitions at 12 and 36 MHz shown in the energy level diagram of Fig. 2 (main text), and the observed decay rate gives a proton $T^*_{2H} = 9.5 \mu$s. Note that the theoretical precision of the transition frequencies measured by ESEEM is much better than that of the fieldswept spectrum, by a factor $\sim T^*_{2H} / T^*_{2e}\approx 4.2\times10^2$. Even without direct ESEEM information about the $^{13}$C transition frequencies, we may combine the $^{1}$H frequencies, fieldswept data and known $^{1}$H hyperfine tensor to obtain a most likely crystal orientation, and then calculate the $^{13}$C parameters from the known carbon hyperfine tensor. \\
\subsection{Details of calculation for observable signal}
The free-evolution Hamiltonian acting on each nuclear spin in the spin-down manifold is of the form
\begin{equation}
\mathcal{H}_{\downarrow j} = \sqrt{(\omega_j - A_j/2)^2 + (B_j/2)^2 }\hspace{1mm} I^j_{\tilde{z_{\downarrow}}} = \omega_{\downarrow j}\hspace{1mm} I^j_{\tilde{z_{\downarrow}}}
\end{equation}
where $I^j_{\tilde{z_{\downarrow}}}$ is the tilted nuclear spin operator along the quantization axis for spin $j$. Evolution of the density matrix terms on the right side of Eq. 5 (main text) under this form of Hamiltonian leads to the time-varying state in Eq. 6 (main text). \\
\indent To perform readout, a selective $\pi/2$ pulse is applied on the electron transition corresponding to the nuclear spin term $\textbf{E}^H_0\otimes \textbf{E}^C_1$. The observable is the $\hat{z}$-magnetization of the electron on this transition, represented by the operator
\begin{equation}
M^{01}_z = \ket{\downarrow}\bra{\downarrow} \otimes \textbf{E}^H_0\otimes \textbf{E}^C_1 -\ket{\uparrow}\bra{\uparrow} \otimes \textbf{E}^H_{0'}\otimes \textbf{E}^C_{1'}
\end{equation}
The normalized signal as function of delay time is therefore
\begin{equation}
S(\tau) = \frac{Tr\left(M^{01}_z\rho(\tau)\right)}{Tr\left((M^{01}_z)^2\right)}  =  \frac{1+ \cos{\left((\omega_{\downarrow C} - \omega_{\downarrow H})\tau\right)}}{2}
\end{equation}
where $\rho(\tau)$ is the state of the system after the second GRAPE pulse. \\
\subsection{Finite bandwidth effects on pulse correction}
\begin{figure*}[ht]
\includegraphics[width= 0.9\columnwidth]{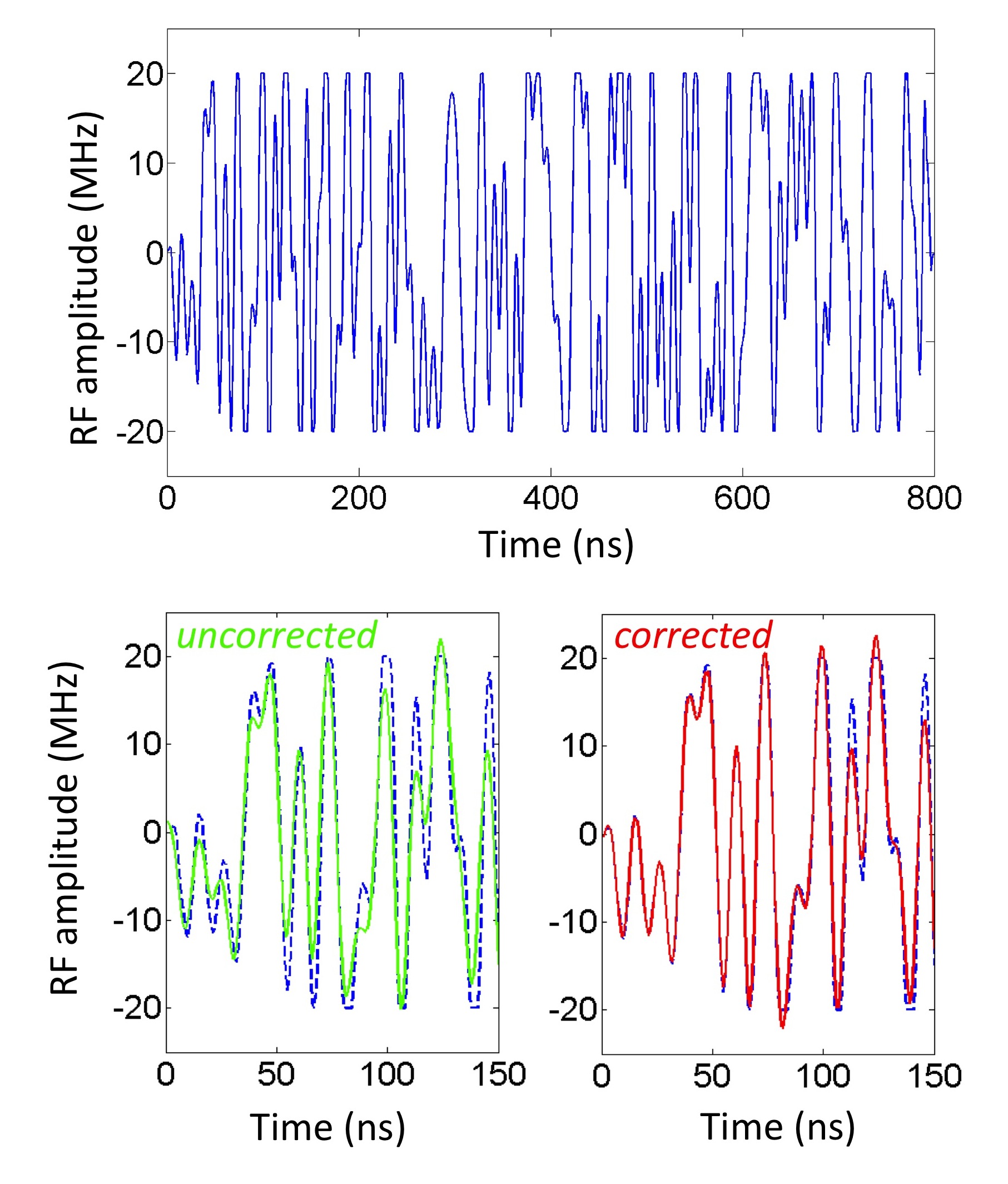} 
\caption{\textbf{top} One quadrature of the envelope waveform for the 800 ns GRAPE pulse used to create double-nuclear coherence. \textbf{lower left} The first 150 ns section of the pulse, with ideal pulse (blue dashed line) and the pulse measured by a pickup loop antenna (green solid line).  \textbf{lower right} The measured pulse (red line) after iterating a feedback/correction protocol, compared with the ideal pulse (blue dashed line). Note that the waveform cannot be perfectly corrected to match the ideal at all points.} \label{grape}
\end{figure*}

\begin{figure*}[ht]
\includegraphics[width= 0.9\columnwidth]{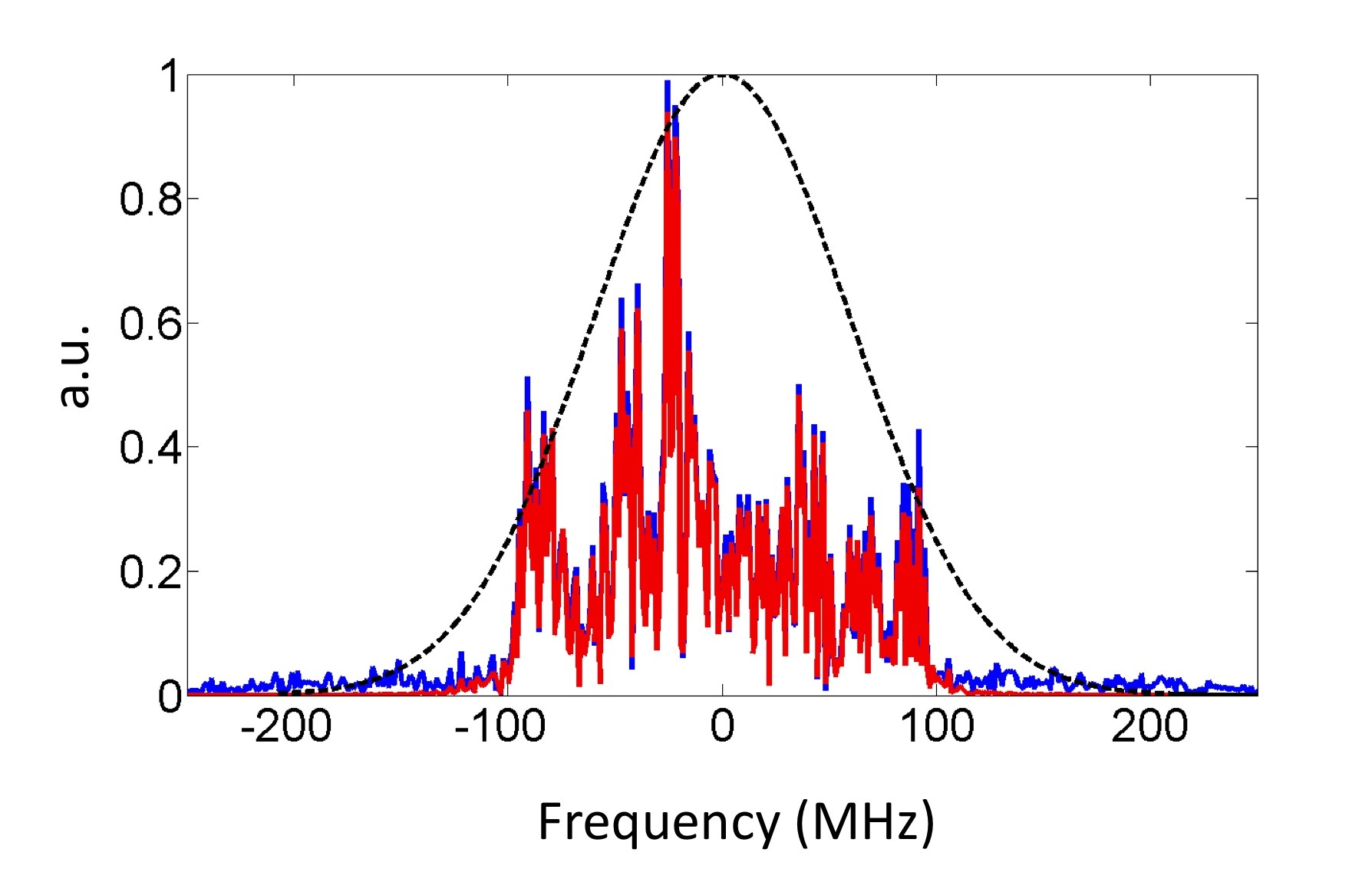} 
\caption{Power spectral density for the pulse in figure~\ref{grape} obtained by taking its Fourier transform. The blue dashed line is the ideal pulse, and the red line corresponds to the measured, corrected pulse. The filter function of the Al loop-gap resonator is shown by the black dashed line.} \label{grapefft}
\end{figure*}

As mentioned in the main text, the finite bandwidth of the resonator, together with imperfections in pulse generation and amplification made it necessary to implement a pulse correction scheme utilizing a pickup loop antenna positioned close to the resonator. The limited resonator bandwidth and microwave power available, however, prevented the pulses from being ideally corrected. Figure~\ref{grape} (top) shows one quadrature of the ideal GRAPE pulse designed to create the 22 MHz nuclear coherence. The lower panels show a section of the ideal pulse compared to the waveform as measured by the loop antenna, before and after our pulse correction scheme is applied. Although the corrected pulse is a better match to the ideal, certain sections involving high frequency components cannot be properly corrected. The resonator filter function is shown in comparison with the power spectrum of the GRAPE pulse in Figure~\ref{grapefft}. We expect that in an improved setup with larger resonator bandwidth and/or narrower system spectral width, experimental fidelities will much more closely match those of simulation.
\end{document}